\documentclass{article}
\usepackage[margin=1in]{geometry}
\usepackage{amsmath,amssymb}
\usepackage{microtype}
\usepackage{xcolor}
\usepackage{array}
\usepackage{booktabs}
\usepackage{titlesec}
\usepackage{hyperref}

\newcolumntype{+}{!{\vrule width 2pt}}

\newlength\savedwidth

\title{\textbf{Machine Learning of the Prime Distribution}}
\author{Alexander Kolpakov\thanks{University of Neuch\^atel, Switzerland; \texttt{kolpakov.alexander@gmail.com}} \and Aidan Rocke\thanks{Solomonoff Consulting, Netherlands}}
\date{}

\titleformat{\section}
{\normalfont\large\bfseries}{\thesection}{1em}{}
\titleformat{\subsection}
{\normalfont\normalsize\bfseries}{\thesubsection}{1em}{}

\begin{document}
\maketitle

\begin{abstract}
In the present work we use maximum entropy methods to derive several theorems in probabilistic number theory, including a version of the Hardy--Ramanujan Theorem. We also provide a theoretical argument explaining the experimental observations of Yang--Hui He about the learnability of primes, and posit that the Erd\H{o}s--Kac law would very unlikely be discovered by current machine learning techniques. Numerical experiments that we perform corroborate our theoretical findings.
\end{abstract}

\section*{Introduction}

Below we briefly recall some known results from Kolmogorov's complexity theory and algorithmic randomness. The reader may find a detailed exposition of this theory in the monographs \cite{grunwald2008algorithmic, li-vitanyi}. Here, however, we assume that most fundamental notions in computability theory are known to the reader. We also provide interpretations for some of these results in the wider epistemological context.  

\subsection*{Kolmogorov's Invariance Theorem}

Let $U$ be a Turing--complete language that is used to simulate a universal Turing machine. Let $p$ be an input of $U$ that produce a given binary string $x \in \{0,1\}^*$. Then the \textit{Kolmogorov Complexity} (or \textit{Minimal Description Length}) of $x$ is defined as

\begin{equation}
    K_U(x) = \min_{p} \{|p|: U\circ p = x\}
\end{equation}
where $U \circ p$ denotes the output of $U$ on input $p$.

Kolmogorov's Invariance Theorem states that the above definition is (on the large scale) invariant of the choice of $U$. Namely, any other Turing--complete language (or, equivalently, another universal Turing machine) $U'$ satisfies

\begin{equation}
\forall x \in \{0,1\}^* : \lvert K_U(x)-K_{U'}(x) \rvert \leq O(1)
\end{equation}

In other terms,

\begin{equation}
\forall x \in \{0,1\}^* : - c(U, U') \leq K_U(x) - K_{U'}(x) \leq c(U, U')
\end{equation}
for some positive constant $c(U, U')$ that depends only on $U$ and $U'$.  

The minimal description $p$ such that $U \circ p = x$ serves as a natural representation of the string $x$ relative to the Turing--complete language $U$. However, it turns out that $p$, and thus $K_U(x)$, is not computable.   

\subsection*{Levin's Universal Distribution} 

The \textit{Algorithmic Probability} of a binary string $x$ can be defined as the probability of $x$ being generated by $U$ on random input $p$, where we consider $p$ to be a binary string generated by fair coin flips:
\begin{equation}
    P(x) = \sum_{p \,:\, U\circ p = x} 2^{-|p|}
\end{equation}

However, this quantity is not well--defined: we can choose one such input $p$ and use it as a prefix for some $p'$ that is about $\log_2 k$ bits longer than $p$ and such that $U$ produces the same binary string: $U\circ p' = x$. Then $2^{-|p'|} \sim 2^{-|p|}/k $, and we have that 
\begin{equation}
    P(x) \geq 2^{-|p|}\, \sum_k \frac{1}{k}
\end{equation}
for $k$ from any subset of integers. Thus, neither $P(x) \leq 1$, nor even has it to be finite. 

Levin's idea effectively formalizes Occam's razor: we need to consider prefix--free Turing--complete languages only. Such languages are easy to imagine: if we agree that all documents end with the instruction \verb=\end{document}= that cannot appear anywhere else, then we have a prefix--free language. 

Given that any prefix--free code satisfies the Kraft-McMillan inequality, we obtain the Universal 
Distribution: 

\begin{equation}
2^{-K_U(x)} \leq P(x) = \sum_{p \,:\, U\circ p = x} 2^{-|p|} \leq 1
\end{equation}
where from now on we consider $U$ to be prefix--free, and $K_U$ is now corresponds to \textit{prefix--free Kolmogorov Complexity}. 

\subsubsection*{Interpretation}

The above facts may also be given a physical interpretation: in a computable Universe, given a phenomenon with encoding $x \in \{0,1\}^*$ generated by a physical process, the probability of that phenomenon is well--defined and equal to the sum over the probabilities of distinct and independent causes. The prefix--free condition is precisely what guarantees causal independence. 

Furthermore, Levin's Universal Distribution formalizes Occam's razor as the most likely cause for that process is provided by the minimal description of $x$ and more lengthy explanations are less probable causes. 

\subsection*{Levin's Coding Theorem}

In the setting of prefix--free Kolmogorov complexity, Levin's Coding theorem states that 

\begin{equation}
K_U(x) = - \log_2 P(x) + O(1)
\end{equation}
or, in other terms, 
\begin{equation}
P(x) = \Theta\left( 2^{-K_U(x)} \right)
\end{equation}

The algorithmic probability of $x$ was defined by Solomonoff simply as 
\begin{equation}
    R(x) = 2^{-K_U(x)} 
\end{equation}
that is proportional to the leading term in the Universal Distribution probability.  

\subsubsection*{Interpretation}

Relative to a prefix--free Turing--complete language $U$ (or, equivalently, a universal prefix--free Turing machine), the number of fair coin flips required to generate the shortest program that outputs $x$ is on the order of $\sim K_U(x)$. 

\subsection*{Maximum Entropy via Occam's razor}

Given a discrete random variable $X$ with computable probability semi--measure $P$, it holds that

\begin{equation}\label{KH}
\mathbb{E}[K_U(X)] = \sum_{x \in X} P(x) \cdot K_U(x) = H(X) + O(1)
\end{equation}
where $H(X)$ is the Shannon Entropy of $X$ in base $2$.  

Moreover, we have that 
\begin{equation}
    \mathbb{E}[R(X)] = 2^{- \mathbb{E}[K_U(X)]} = 2^{- H(X) + O(1)}
\end{equation}
which means that the expected Solomonoff probability is inverse proportional to the effective alphabet size. 

\subsubsection*{Interpretation}

The Shannon Entropy of a random variable in base $2$ equals its Expected Kolmogorov Complexity up to a constant that becomes negligible in asymptotic analysis. This provides us with a precise answer to von Neumann's original question. 

As follows from \eqref{KH}, the average Kolmogorov Complexity and Entropy have the same order of magnitude. Hence machine learning systems that minimise the $KL$--Divergence are implicitly applying Occam's razor.   

However, what exactly do we mean by random variable? In a computable Universe the sample space of a random variable $X$ represents the state--space of a Turing Machine with unknown dynamics whose output sequence is computable. As the generated sequence is computable, it is finite--state incompressible in the worst-case i.e. a normal number. Hence, a random variable corresponds to a stochastic source that is finite--state random. 

This definition comes from the well--known correspondence between finite--state machines and normal numbers that establishes that a sequence is normal if and only if there is no finite--state machine that accepts it. 

\subsection*{Algorithmic Randomness}

Let $x$ be a binary sequence, and $x|_N$ be the initial segment of $N$ bits. Then $x$ is \textit{algorithmically random} if
\begin{equation}
    K_U(x|_N) \geq N - c 
\end{equation}
for some constant $c \geq 0$. Such a string $x$ is also called \textit{incompressible}. 

Thus, an integer $n \in \mathbb{N}$ is algorithmically random (or incompressible) if  
\begin{equation}
    K_U(n) = \log_2 n + O(1) 
\end{equation}
and a discrete random variable $X$ taking $N$ values with computable distribution is algorithmically random if 
\begin{equation}
    \mathbb{E}[K_U(X)] = \log_2 N + O(1)
\end{equation}
which corresponds to the maximal entropy case of the uniform distribution. 

\subsubsection*{Interpretation}

A random binary string cannot be compressed because it does not contain any information about itself. Thus, a random number $n$ needs $\log_2 n$ bits to be written in its binary form, and a random variable is ``truly random'' in a computable Universe if its expected Kolmogorov complexity is maximal.

\section*{Maximum Entropy Methods for Probabilistic Number Theory}

Below we provide an illustration of information--theoretical approach to classical theorems in number theory. Some proofs using the notions of entropy and Kolmogorov's complexity have already been known \cite{kontoyiannis2008counting}, and other proofs below are definitely new. We prefer to keep both kinds of them to retain the whole picture.  

\subsection*{The Erd\H{o}s--Euclid theorem}

This information--theoretic adaptation of Erd\H{o}s' proof of Euclid's theorem is originally due to Ioannis Kontoyiannis \cite{kontoyiannis2008counting}. In essence, this proof demonstrates that the information content of finitely many primes is insufficient to generate all the integers.

Let $\pi(N)$ be the number of primes that are less or equal to a given natural number $N$. Let us suppose that the set of primes $\mathbb{P}$ is finite so we have $\mathbb{P}=\{p_i\}_{i=1}^{\pi(N)}$ where $\pi(N)$ is constant for $N$ big enough. 

Then we can define a uniform integer--valued random variable $Z$ chosen uniformly from $[1,N]$, such that 
\begin{equation}
Z = \left(\prod_{i=1}^{\pi(N)} p_i^{X_i}\right) \cdot Y^2 
\end{equation}
for some integer--valued random variables $1 \leq Y \leq N$ and $X_i \in \{0,1\}$, such that $Z/Y^2$ is square--free. In particular, we have that $Y \leq \sqrt{N}$, thus the upper bound for Shannon's Entropy from Jensen's inequality implies:

\begin{equation}
H(Y) \leq \log_2 \sqrt{N} = \frac{1}{2} \log_2 N 
\end{equation}

Also, since $X_i$ is a binary variable, we have $H(X_i) \leq 1$. 

Then, we compute
\begin{equation}
H(Z) = \log_2 N	
\end{equation}

Moreover, we readily obtain the following inequality:
\begin{equation}
\begin{split}
    &H(Z) = H\left(Y, X_1, \ldots, X_{\pi(N)}\right)  \\
    &\leq H(Y) + \sum^{\pi(N)}_{i=1} H(X_i) \\
    &\leq \frac{1}{2} \log_2 N + \pi(N) 
\end{split}
\end{equation}

which implies

\begin{equation}
\pi(N) \geq \frac{1}{2} \log_2 N 
\end{equation}

This clearly contradicts the assumption that $\pi(N)$ is a constant for any natural $N$, and provides us with a simple but far from reasonable lower bound.

\subsection*{Chebyshev's theorem via Algorithmic Probability}

An information--theoretic derivation of Chebyshev's Theorem \cite{chebychev1852sur}, an important precursor of the Prime Number Theorem, from the Maximum Entropy Principle. Another proof was given by Ioannis Kontoyiannis in \cite{kontoyiannis2008counting}.

Chebyshev's Theorem is a classical result that states that

\begin{equation}
\sum_{p \leq N} \frac{1}{p} \cdot \ln p \sim \ln N 
\end{equation}
where we sum over the primes $p \leq N$. 

Here, two functions $f(n) \sim g(n)$ are asymptotically equivalent once $\lim_{n\to \infty} \frac{f(n)}{g(n)} = 1$. 

In information--theoretical terms, the expected information gained from observing a prime number in the interval $[1,N]$ is on the order of $\sim \log_2 N$. 

For an integer $Z$ sampled uniformly from the interval $[1,N]$ we may define its random prime factorization in terms of the random variables $X_p$: 
\begin{equation}
Z = \prod_{p \leq N} p^{X_p} 
\end{equation}

As we have no prior information about $Z$, it has the maximum entropy distribution among all possible distributions on $[1,N]$.

Since the set of finite strings is countable so there is a $1$--to--$1$ map from $\{0,1\}^*$ to $\mathbb{N}$ and we may define the Kolmogorov Complexity as a map from integers to integers, $K_U: \mathbb{N} \rightarrow \mathbb{N}$. As almost all finite strings are incompressible \cite[Theorem 2.2.1]{li-vitanyi}, it follows that almost all integers are algorithmically random. Thus, for large $N$ and for $n \in [1,N]$, we have $K_U(n) = \log_2 n + O(1)$ almost surely. 

Thus,

\begin{equation}
\begin{split}
&\mathbb{E}[K_U(Z)] \sim \mathbb{E}[\log_2 Z] = \\
&\frac{1}{N} \sum_{k=1}^N \log_2 k = \frac{\log_2(N!)}{N} \sim \log_2 N 
\end{split}
\end{equation}

by using Stirling's approximation

\begin{equation}
    \log_2(N!) = N \log_2 N - N \log_2 e + O(\log_2 N)
\end{equation}

On the other hand, 
\begin{equation}
    \mathbb{E}[K_U(Z)] \sim \mathbb{E}[\log_2 Z] = \sum_{p \leq N} \mathbb{E}[X_p] \cdot \log_2 p
\end{equation}

All of the above implies 

\begin{equation}
\mathbb{E}[K_U(Z)] \sim \sum_{p \leq N} \mathbb{E}[X_p] \cdot \log_2 p \sim \log_2 N
\end{equation}

Here, we may apply the Maximum Entropy Principle to $X_p$, provided that $\mathbb{E}[X_p]$ is fixed: then $X_p$ belongs to the geometric distribution \cite[Theorem 12.1.1]{cover-thomas}. This can also be verified directly as shown in \cite{kontoyiannis2008counting}. 

Thus, we compute: 

\begin{equation}
\mathbb{E}[X_p] = \sum_{k \geq 1} P(X_p \geq k) = \sum_{k \geq 1} \frac{1}{N} \left\lfloor \frac{N}{p^k} \right\rfloor \sim \frac{1}{p}
\end{equation}

Thus, we rediscover Chebyshev's theorem: 

\begin{equation}
\begin{split}
&H(X_{p_1}, \ldots, X_{p_{\pi(N)}}) = H(Z) \sim \mathbb{E}[K_U(Z)] \\
&\sim \sum_{p \leq N} \frac{1}{p} \cdot \log_2 p \sim \log_2 N 
\end{split}
\end{equation}

It should be noted that the asymptotic equivalence above does not depended on the logarithm's base (as long as the same base is used on both sides). 

\subsection*{The Hardy--Ramanujan Theorem from Information Theory}

Based on the ideas explained in the previous paragraphs, we can deduce a version of the classical Hardy--Ramanujan theorem. 

\subsubsection*{The Probability of a Prime Factor}

Given an integer $Z$ chosen uniformly from $[1,N]$ with prime factorisation
\begin{equation}
Z = \prod_{p \leq N} p^{X_p}
\end{equation}
we observe that 
\begin{equation}
\begin{split}
    &P(X_1 \geq k_1, \ldots, X_n \geq k_n) = \frac{1}{N} \left\lfloor \frac{N}{p_1^{k_1} \ldots p_n^{k_n}} \right\rfloor \sim \\
    &{\prod}_{i\leq n} \frac{1}{N} \left\lfloor \frac{N}{p_i^{k_i}} \right\rfloor = \prod_{i\leq n} P(X_i \geq k_i) 
\end{split}
\end{equation}
which means that $X_i$'s are asymptotically independent (i.e. independent in the large $N$ limit). 

In particular, the probability of having $p \in \mathbb{P}$ as a prime factor satisfies

\begin{equation}
P(X_p \geq 1) \sim \frac{1}{p}
\end{equation}

Likewise, the probability that two distinct primes $p,q \in \mathbb{P}$ are happen as factors simultaneously satisfies

\begin{equation}
P(X_p \geq 1 \land X_q \geq 1) \sim \frac{1}{p} \cdot \frac{1}{q}
\end{equation}

Thus, any two sufficiently large prime numbers $p, q \in \mathbb{P}$ are statistically independent. 

\subsubsection*{The Expected Number of Unique Prime Factors}

For any integer $Z$ sampled uniformly from $[1,N]$, we may define its number of unique prime factors $w(Z) = \sum_{p \leq N} I(X_p \geq 1)$. Thus, we calculate the expected value 

\begin{equation}
\mathbb{E}[w(Z)] = \sum_{p \leq N} P(X_p \geq 1) \sim \sum_{p \leq N} \frac{1}{p} = \ln \ln N + O(1)
\end{equation}
where we use Mertens' $2$nd Theorem for the last equality. 

\subsubsection*{The Standard Deviation of $w(Z)$}

Given our previous analysis, the random variables $I_p = I(X_p \geq 1)$ are asymptotically independent, so that the variance of $w(Z)$ satisfies 

\begin{equation}
\begin{split}
&\textrm{Var}[w(Z)] \sim \sum_{p \leq N} \mathbb{E}[I_p^2] - \mathbb{E}[I_p]^2 \sim \\
&\sum_{p \leq N} \big(\frac{1}{p} - \frac{1}{p^2} \big) = \ln \ln N + O(1) 
\end{split}
\end{equation}
since $\sum_{p \leq N} \frac{1}{p^2} \leq \frac{\pi^2}{6}$.

\subsubsection*{The Hardy--Ramanujan Theorem}

For any positive $k > 0$ and a random variable $Y$ with $\mathbb{E}[Y] = \mu$ and $\textrm{Var}[Y] = \sigma^2$, Chebyshev's inequality informs us that
\begin{equation}
    P(|Y - \mu| > k \sigma) \leq \frac{1}{k^2}
\end{equation}

With $Y = w(Z)$ and $k = (\ln \ln N)^{\varepsilon}$ we obtain a probabilistic version of the Hardy--Ramanujan theorem. Namely, for a random integer $Z \sim \mathcal{U}([1,N])$, and any function $\psi(N) \sim (\ln \ln N)^{1/2 + \varepsilon}$, in the large $N$ limit we have 
\begin{equation}
    | w(Z) - \ln \ln N | \leq \psi(N) 
\end{equation}
with probability 
\begin{equation}
    1 - O\left(\frac{1}{(\ln \ln N)^{2 \varepsilon}}\right)
\end{equation}

\subsection*{Discussion and Conclusions}

The Erd\H{o}s--Kac theorem states that
\begin{equation}
\frac{\omega(Z)- \ln \ln N}{\sqrt{\ln \ln N}} 
\end{equation}
converges to the standard normal distribution $\mathcal{N}(0,1)$ as $N \rightarrow \infty$.

This theorem is of great interest to the broader mathematical community as it 
is impossible to guess from empirical observations. In fact, it is far from certain that Erd\H{o}s and Kac would have proved the Erd\H{o}s--Kac theorem if its precursor, the Hardy--Ramanujan theorem, was not first discovered. 

More generally, in the times of Big Data this theorem forces the issue of determining how some scientists were able to formulate correct theories based on virtually zero empirical evidence. 

In our computational experiments $Z$ was chosen to run through all possible $N$--bit integers with $N=24$, and no normal distribution emerged according to the D'Agostino--Pearson test. The $p$--value associated to this test equals the probability of sampling normally distributed data that produces at least as extreme value of the D'Agostino--Pearson statistics as the actual observations. Thus, if the $p$--value is small it may be taken as evidence against the normality hypothesis. In our case we obtained a $p$--value of $0.0$. 

For comparison, the Central Limit Theorem clearly manifests itself for $2^{16}$, not even $2^{24}$, binomial distribution samples. In this case, the $p$--value associated to the D'Agostino--Pearson test equals $\sim 0.4657$. 

The code used in our numerical experiments is available on GitHub \cite{github}, and all computations can be reproduced on a laptop computer such as MacBook Pro M1 with 8~GB RAM, or comparable. 

In order to observe the normal order in the Erd\H{o}s--Kac law, one would need $Z$ to reach about $2^{240}$, not $2^{24}$, instead \cite{renyi-turan-1958}. 

Thus, non--trivial scientific discoveries of this kind that are provably beyond the scope of computational induction (and hence machine learning) do not yet have an adequate explanation. 

\section*{Learning the Prime Distribution}

Below we provide a theoretical study of the Kolmogorov complexity of the prime distribution. 

\subsection*{Entropy of Primes}

In what follows $p \in \mathbb{P}$ will always be a fixed prime number, and $X \in [1, N]$ will be chosen at random among the first $N$ natural numbers. Considering prime numbers as random variables, or as elements of random walks, goes back to Billingsley's heuristics in \cite{billingsley1973prime}. 

From the information--theoretic perspective, Chebyshev's theorem states that the average code length of $X$ expressed as the sequence of prime exponents $X_{p_1}, \ldots, X_{p_{\pi(N)}}$ satisfies

\begin{equation}
H(X) = H(X_{p_1}, \ldots, X_{p_{\pi(N)}}) \sim \sum_{p\leq N} \frac{1}{p}\cdot \ln p \sim \ln N 	
\end{equation}
where we use the natural logarithm entropy instead of binary entropy. As previously noted, this is only a matter of computational convenience. 

It turns out that the entropy of $X$, which is almost surely composite in the large $N$ limit, essentially depends on the available primes $p \leq N$.  

A given non--negative integer $n$ has binary code length 
\begin{equation}
    \ell(n) = \lfloor \log_2 n \rfloor + 1 \sim \log_2 n + O(1)
\end{equation}

Given an integer $n \in [1,N]$, we need at most $\ell(n) = $ bits to encode it, and thus it can be produced by a program of length $\leq \ell(n)$. Note that $\ell(N)$ is so far irrelevant here as we need a prefix--free encoding and do not consider adding zero bits for padding all binary strings to same length.

By Levin's coding theorem, 
\begin{equation}
    P(X = n) = \Theta\left( 2^{-K_U(n)} \right)
\end{equation}
 and thus we have  
\begin{equation}
    2^{-\ell(n)} \leq P(X = n) \leq \alpha \cdot 2^{-\ell(n)}
\end{equation}
with $\alpha > 1$, for most numbers $n\in[1,N]$, as most binary strings are incompressible in the large $N$ limit \cite[Theorem 2.2.1]{li-vitanyi}. 

Thus, we may as well resort to the following Ansatz:
\begin{equation}
    P(X = n) = \frac{1}{n}
\end{equation}
as it provides a computable measure on $[1,N]$ that is roughly equivalent to the initial Levin's distribution. The same discrete measure arises in the heuristic derivation of Benford's law \cite{wolfram_benfords_law}. 

Let us consider a random variable $Y$ that generates \textit{primes} within $[1, N]$ with probability 
\begin{equation*}
    P(Y = p) = \frac{1}{p} 
\end{equation*}
that appears to represent the \textit{algorithmic} probability of a prime rather than its frequentist probability. 

Then we can write 
\begin{equation*}
    H(Y) = - \sum_{p\leq N} P(Y=p) \cdot \ln P(Y=p) = \sum_{p\leq N} \frac{\ln p}{p} \sim \ln N
\end{equation*}

There are $\pi(N)$ primes $p$ satisfying $p \leq N$, and thus we need exactly $\pi(N)$ random variables $Y_i$ to generate them all. Each $Y_i$ has the same distribution as $Y$, and we \textit{assume} that $Y_i$'s are independent. 

Let $\widetilde{Y}_N = (Y_1, \ldots, Y_{\pi(N)})$ be the ordered sequence generating all primes in $[1,N]$. Then Shannon's source coding theorem informs us that $\widetilde{Y}_N$ needs no less than $\pi(N)\cdot H(Y)$ bits to be encoded without information loss. Any smaller encoding will almost surely lead to information loss. This means that
\begin{equation}
    H(\widetilde{Y}_N) = \pi(N)\cdot H(Y) + O(1)
\end{equation}
Thus,
\begin{equation}
    \mathbb{E}[K_U(\widetilde{Y}_N)] = H(\widetilde{Y}_N) + O(1) = \pi(N)\cdot H(Y) + O(1) \sim \pi(N)\cdot \ln(N) \sim N
\end{equation}
as the Prime Number Theorem provides the asymptotic equivalence
\begin{equation}
    \pi(N) \sim \frac{N}{\ln N}
\end{equation}

On the other hand, the most obvious encoding of primes is the $N$--bit string where we put $1$ in position $k$ if $k$ is prime, and $0$ if $k$ is composite. The above equality for expected Kolmogorov's complexity of $\widetilde{Y}_N$ implies that for large values of $N$ this string is almost surely incompressible.

\subsection*{Discussion and Conclusions}

The above discussion provides a theoretical corroboration of the experimental fact observed by Yang--Hui He in \cite{he}. Namely, the complexity of ``machine learning'' the prime distribution on the interval $[1, N]$ is equivalent to learning an algorithmically random sequence. Thus, the true positive rate of any model predicting primes should be very low. 

There are, however, several theoretical issues with the above argument. One is that the proposed Benford's probability is not a finite measure, as we have
\begin{equation}
    \sum_{k \leq N} P(X = k) = \sum_{k\leq N} \frac{1}{k} = \ln N + O(1)
\end{equation}

The $1$st Mertens' theorem gives that summing over the primes only gives 
\begin{equation}
    \sum_{p \leq N} P(Y = p) = \sum_{p\leq N} \frac{1}{p} = \ln \ln N + O(1)
\end{equation}
which is a much smaller quantity asymptotically, yet unbounded. 

We can apply some regularization, such as setting 
\begin{equation}
    P(X = n) = \frac{1}{n^s}
\end{equation}
for some $s > 1$, and thus obtaining a finite measure and trying to study its $s \to 1$ limit. This measure is not normalized to be a probability measure, however this is not very crucial (e.g. Levin's universal probability is not normalized either). 

Indeed, we shall have
\begin{equation}
    \sum_{k=1}^{\infty} P(X = k) = \zeta(s) < \infty
\end{equation}
for any $s > 1$. 

Moreover,
\begin{equation}
    H_s(Y)  = s \cdot \sum_{p\leq N} \frac{\ln p}{p^s} \xrightarrow[s \to 1]{} H(Y)
\end{equation}

In contrast, $P$ will not converge to a finite measure as $\zeta(s)$ has a pole at $s=1$. 

Given the fact that the Prime Encoding produces a bit--string that is algorithmically random (at least, asymptotically), the expected True Positive Rate (TPR) for inferring the first $N$ numbers as primes or composites is on the order of
\begin{equation}
    \mathbb{E}[TPR] \sim \frac{1}{N} \sum_{k=1}^N P(X=k) = \frac{1}{N} \sum^N_{k=1} \frac{1}{n} = \frac{\ln N}{N}
\end{equation}
and thus tends to $0$ as $N$ becomes large. Our numerical experiments corroborate the fact that the TPR of a machine learning model is indeed small and, moreover, that it diminishes as $N$ grows. 

Another point worthy discussion is the independence of primes $Y_1, \ldots, Y_{\pi(N)}$. This assumption allows Shannon's source coding theorem into play, however it does not seem to fully hold (even theoretically). Indeed, a Turing machine that enumerates any finite initial segment of primes exists. Arguably, since $K(n) \sim log_2 n$ for \textit{most} numbers, we may write the following upper bound for the complexity of primes up to $2^N$
\begin{equation}
    K(\pi(2^N)) + O(1) \sim \log_2\left( \frac{2^N}{N \ln 2} \right) + O(1) \sim N - \log_2 N + O(1)
\end{equation}

Thus, the primes are somewhat compressible, as their complexity is not outright $N - O(1)$. However, our result about the expected Kolmogorov complexity of primes holds in the sense of asymptotic equivalence, and $N - \log_2 N \sim N$ in the large $N$ limit. 

\subsection*{Numerical Experiments}

We posit that a machine learning model may not be reliably used to predict the locations of prime numbers. Indeed, $X$ being algorithmically random means that no prediction can be reasonably inferred for it by using inductive learning. 

Previous experiments on prime number inference using deep learning were done in \cite{he}, and showed a very low true positive rate $(\sim 10^{-3})$. The neural network had a three--layer architecture, and no specific training was performed. Modern tree--based classifiers are approximating the Kolmogorov complexity very efficiently by using Huffman encoding or a more advanced variant of thereof. 

For example, XGBoost often outperforms other models in Kaggle competitions, especially on tabular data and in classification tasks. Thus, we may take XGBoost as a more practical experimental model. 

We performed XGBoost experiments on prime learning for $N$--bit integers with $N=18$ and $N=24$. Each integer was represented as a binary string of length $N$ with leading zeroes where appropriate. The code used for our experiments is available on GitHub \cite{github}, and the computations may be reproduced on a GPU--equipped laptop, e.g. MacBook Pro M1 with 8 GB memory. 

For $N=18$, there are $23'000$ primes and $239'144$ composites out of $262'144$ numbers in total. We have the following probability confusion matrix:
\begin{equation*}
\begin{array}{c|cc}
{} & Composite & Prime \\ \hline
Composite & 0.638886 & 0.273278 \\
Prime & 0.058209 & 0.029628
\end{array}
\end{equation*}

%[[0.638886 0.273278]
%[0.058209 0.029628]]

For $N=24$, there are $1'077'871$ primes and $15'699'345$ composites out of $16'777'216$ numbers in total. The probability confusion matrix turns out 
\begin{equation*}
\begin{array}{c|cc}
{} & Composite & Prime \\ \hline
Composite & 0.635707 & 0.300054 \\
Prime & 0.041912 & 0.022326
\end{array}
\end{equation*}

%[[0.635707 0.300054]
% [0.041912 0.022326]]

In both cases we used Bayesian hyper--parameter optimization with HyperOpt, and 80\% / 20\% train/test split. This achieves a better (by an order of magnitude) true positive rate than in \cite{he}, which is still insignificant. In fact, as follows from the above confusion matrices, the true positive rate declines as $N$ grows. All this is expected given our theoretical analysis, and our experiments corroborate the theoretical conclusion that ``machine learning'' primes turns out no better than guessing a random number. 

As shown in \cite{blake2023}, the recent experiments with Neural Network classifiers for \textit{semi--primes} largely fail to infer the semi--prime distribution. More precisely, for $N=426$ detecting semi--primes with $N$--bit factors has the number of false negatives on par with the number of true positives (both on the order of $\sim 0.25$). It is worth mentioning that the number of false positives ($\sim 0.05$) is relatively small. However, the problem considered in \cite{blake2023} is substantially different from ours. 

\subsection*{Data accessibility} All numerical experiments can be reproduced by accessing the associated GitHub repository \cite{github}.

\subsection*{Acknowledgements} We would like to thank Anders S\"odergren, Ioannis Kontoyiannis, Hector Zenil, Steve Brunton, Marcus Hutter, Cristian Calude, Igor Rivin, and the PLOS manuscript reviewers for their constructive feedback in the preparation of this manuscript. 

A.K. was supported by the Swiss National Science Foundation project PP00P2--202667. A.R. thanks the University of Neuch\^atel for hospitality during his research stay in July 2023.

%\subsection*{Funding} A.K. was supported by the Swiss National Science Foundation project PP00P2--202667. A.R. thanks the University of Neuch\^atel for hospitality during his research stay in July 2023. 

%\subsection*{Ethics statement} This work did not involve any active collection of human data.

%\subsection*{Data accessibility} All numerical experiments can be reproduced by accessing the associated GitHub repository \cite{github}.

%\subsection*{Competing interests statement} The authors declare no competing interests.

%\subsection*{Authors’ contributions} A.K. and A.R. conceived the mathematical models, interpreted the computational results, and wrote the paper. A.K. implemented and performed most of the numerical experiments. Both authors gave final approval for publication.

%\nolinenumbers

% Either type in your references using
% \begin{thebibliography}{}
% \bibitem{}
% Text
% \end{thebibliography}
%
% or
%
% Compile your BiBTeX database using our plos2015.bst
% style file and paste the contents of your .bbl file
% here. See http://journals.plos.org/plosone/s/latex for 
% step-by-step instructions.
% 

\end{document}